\documentclass[12pt]{article} 
\usepackage{amsmath} 
\usepackage{amssymb}
\usepackage{setspace}
\usepackage{graphicx}
\usepackage{hyperref}

\numberwithin{equation}{section}

\newcommand{\qed}{\nobreak \ifvmode \relax \else
      \ifdim\lastskip<1\cdot5em \hskip-\lastskip
      \hskip1\cdot5em plus0em minus0\cdot5em \fi \nobreak
      \vrule height0\cdot75em width0\cdot5em depth0\cdot25em\fi}

\begin{document}

\title{State Vector Determination By a Single Tracking Satellite} 

\author{Kerry M. Soileau} 
\date{April 27, 2004}

\maketitle

\begin{abstract}
Using only a single tracker capable only of range measurements to an orbiting object in an unknown Keplerian orbit, it is theoretically possible to narrow the possibilities for the tracker's state vector to at worst a few, and at best only two.
\end{abstract}


\doublespacing

\section{Introduction}
Satellite orbit determination using ranging from a fixed ground station is well-understood, see for example \cite{erwin06}. In this article we explore the limits of orbit determination in case the ranging station is not fixed to the surface of a planet, but is itself in an accurately-known circular orbit.

\section{Preliminaries}
Suppose we have a satellite in circular orbit (henceforward called the \textbf{tracker}.) We assume the tracker is capable of measuring range and range rate to another satellite in an unknown  orbit (henceforward called the \textbf{trackee}.). We denote the time-varying position vectors of the tracker and trackee by $\vec{R}$ and $\vec{r},$ respectively. The only force on the tracker and trackee is assumed to be that due to a large uniformly dense body. To simplify the following derivation, we choose our time unit to be the time taken for the tracker to travel one radian in its orbit, and our distance unit to be the radius of the tracker's orbit. We may then write $\vec{R}=\hat{R}$ and $\vec{r}=r \, \hat{r}$ where $r=|\vec{r}\,|>0$ for all $t.$

In the following derivation, we will make use of the following equations:

\begin{equation}
\hat{R} \cdot \hat{T}=0
\end{equation}

\begin{equation}
\frac{d\hat{R}}{dt}=\hat{T}
\end{equation}

\begin{equation}
\frac{d\hat{T}}{dt}=-\hat{R}
\end{equation}

\begin{equation}
\frac{d\vec{r}}{dt}=\vec{v}
\end{equation}

\begin{equation}
\frac{d\vec{v}}{dt}=-\frac{1}{\left(\vec{r} \cdot \vec{r}\right)^\frac{3}{2}} \vec{r}
\end{equation}

where $\hat{T}$ is the velocity unit vector for the tracker.

\section{Range Squared}

We define $q(t) \equiv |\vec{r}(t)-\hat{R}(t)|^2\cdot$ Then

\begin{equation}
\noindent q(t)=1-2  \vec{r}(t) \cdot  \hat{R}(t)+ \vec{r}(t) \cdot  \vec{r}(t)
\end{equation}

\begin{equation}
\noindent q'(t)=-2 ( \vec{r}(t) \cdot  \hat{T}(t)- \vec{r}(t) \cdot  \vec{v}(t)+ \vec{v}(t) \cdot  \hat{R}(t))
\end{equation}

\begin{equation}
\begin{split}
\noindent q''(t)
=2 \vec{r}(t) \cdot  \hat{R}(t) \left(1+\frac{1}{( \vec{r}(t) \cdot  \vec{r}(t))^{3/2}}\right)-\frac{2}{\sqrt{ \vec{r}(t) \cdot  \vec{r}(t)}}\\
-4
 \vec{v}(t) \cdot  \hat{T}(t)+ 2\vec{v}(t) \cdot  \vec{v}(t)
\end{split}
\end{equation}

\begin{equation}
\begin{split}
\noindent q'''(t)=-\frac{6 \vec{r}(t) \cdot \hat{R}(t) \vec{r}(t) \cdot \vec{v}(t)}{(\vec{r}(t) \cdot \vec{r}(t))^{5/2}}+\frac{2
   \vec{v}(t) \cdot \hat{R}(t)}{(\vec{r}(t) \cdot \vec{r}(t))^{3/2}}\\
   +6 \vec{v}(t) \cdot \hat{R}(t)+\frac{6
   \vec{r}(t) \cdot \hat{T}(t)}{(\vec{r}(t) \cdot \vec{r}(t))^{3/2}}+2 \vec{r}(t) \cdot \hat{T}(t)-\frac{2
   \vec{r}(t) \cdot \vec{v}(t)}{(\vec{r}(t) \cdot \vec{r}(t))^{3/2}} 
\end{split}
\end{equation}

Fix a time $t_0$. In the following we will consider what information about the trackee's orbit can be inferred from the values of the time derivatives of $q$ of orders zero through six at time $t_0$. $\vec{r}$ will mean $\vec{r}(t_0)$, $\vec{v}$ will mean $\vec{v}(t_0)$, $r$ will mean $\left| \vec{r}(t_0) \right|$ and so on.

\subsection{Case $r \neq 1$}
Solving the system

\begin{equation}
\begin{array}{c}
  q(t_0)=m_0 \\
  q'(t_0)=m_1 \\
  q''(t_0)=m_2 \\
  q'''(t_0)=m_3 \\
\end{array}
\end{equation}

we get

\begin{equation}
\vec{r} \cdot \hat{R} \equiv \frac{1}{2} \left(-m_0+r^2+1\right)
\end{equation}

\begin{equation}
\vec{r} \cdot \hat{T} \equiv -\frac{3 { \vec{r} \cdot \vec{v} } \left(-m_0-2 r^5+r^2+1\right)+m_1 \left(3
   r^5+r^2\right)+m_3 r^5}{4 r^2 \left(r^3-1\right)}
\end{equation}

\begin{equation}
\vec{v} \cdot \hat{R} \equiv \frac{-{ \vec{r} \cdot \vec{v} } \left(3 m_0+2 r^5+r^2-3\right)+m_1 \left(r^3+3\right)
   r^2+m_3 r^5}{4 r^2 \left(r^3-1\right)}
\end{equation}

\begin{equation}
\vec{v} \cdot \hat{T} \equiv \frac{-m_0 \left(r^3+1\right)+r^3 \left(-m_2+2 v^2+1\right)+r^5-r^2+1}{4 r^3}	
\end{equation}

\subsubsection{$\vec{v} \cdot \vec{v}$}
Using the condition $q^{(4)}(t_0)=m_4$ and the previous equations, we can determine the possible values for $\vec{v} \cdot \vec{v}$ in terms of $r$ and $\vec{r} \cdot \vec{v}$:

\begin{equation}
\vec{v} \cdot \vec{v}=\frac{\alpha}{\beta}
\end{equation}

where

\begin{equation}
\begin{split}
\alpha=\left(9 r^5+(15
   m_0-15) r^3-6 m_0+6\right) \left( \vec{r} \cdot \vec{v} \right)^2\\
   + \left( (-15 m_1-3 m_3)r^5+3 m_1
   r^2\right)\vec{r} \cdot \vec{v}\\
   -r^{12}
   +r^{10} (m_0+2 m_2+m_4-1)
+ (-3 m_0-m_4+3)r^7\\
+7 r^9
-8 r^6
+ (6 m_0-2 m_2-6)r^4+2 r^3+(4-4 m_0) r
\end{split}
\end{equation}

and 

\begin{equation}
\beta=r^2 \left(r^3-1\right) \left(3 m_0+4 r^5-r^2-3\right)
\end{equation}

\subsubsection{$p_1 \left(r,\vec{r} \cdot \vec{v} \right)$}
The condition $q^{(5)}(t_0)=m_5$ and the previous equations imply the following polynomial equation in $r$ and $\vec{r} \cdot \vec{v}$:

\begin{equation}
p_1\left(r,\vec{r} \cdot \vec{v}\right)=0	
\end{equation}
where
\begin{equation}
p_1\left(r,\vec{r} \cdot \vec{v}\right)=	a_0(r)+a_1(r) \vec{r} \cdot \vec{v}+a_2(r)\left(\vec{r} \cdot \vec{v}\right)^2+a_3(r)\left(\vec{r} \cdot \vec{v}\right)^3
\end{equation}
and

$
a_0(r) \equiv 15 m_0 m_1 r^{15}-48 m_0 m_1 r^{12}+54 m_0 m_1
   r^9-24 m_0 m_1 r^6+3 m_0 m_1 r^3-3 m_0
   m_3 r^{15}-3 m_0 m_3 r^{12}+15 m_0 m_3 r^9-9
   m_0 m_3 r^6-3 m_0 m_5 r^{15}+6 m_0 m_5
   r^{12}-3 m_0 m_5 r^9+36 m_1 m_2 r^{15}-12 m_1
   m_2 r^{12}-36 m_1 m_2 r^9+12 m_1 m_2 r^6+18
   m_1 m_4 r^{15}-24 m_1 m_4 r^{12}+6 m_1 m_4
   r^9-4 m_1 r^{20}-m_1 r^{17}-15 m_1 r^{15}+32 m_1
   r^{14}+48 m_1 r^{12}-50 m_1 r^{11}-54 m_1 r^9+28
   m_1 r^8+24 m_1 r^6-5 m_1 r^5-3 m_1 r^3+6 m_2
   m_3 r^{15}-6 m_2 m_3 r^9+3 m_3 m_4 r^{15}-3
   m_3 m_4 r^{12}-8 m_3 r^{20}+7 m_3 r^{17}+3 m_3
   r^{15}+15 m_3 r^{14}+3 m_3 r^{12}-19 m_3 r^{11}-15
   m_3 r^9+5 m_3 r^8+9 m_3 r^6-4 m_5 r^{20}+9
   m_5 r^{17}+3 m_5 r^{15}-6 m_5 r^{14}-6 m_5
   r^{12}+m_5 r^{11}+3 m_5 r^9
$ 

$  
a_1(r) \equiv -45 m_0^2 r^{13}+144 m_0^2 r^{10}-171 m_0^2 r^7+90 m_0^2
   r^4-18 m_0^2 r-90 m_0 m_2 r^{13}+117 m_0 m_2
   r^{10}-27 m_0 m_2 r^4-45 m_0 m_4 r^{13}+81 m_0
   m_4 r^{10}-36 m_0 m_4 r^7+12 m_0 r^{15}+90 m_0
   r^{13}-72 m_0 r^{12}-288 m_0 r^{10}+126 m_0 r^9+342
   m_0 r^7-84 m_0 r^6-180 m_0 r^4+18 m_0 r^3+36
   m_0 r-270 m_1^2 r^{10}+144 m_1^2 r^7-18 m_1^2
   r^4-99 m_1 m_3 r^{10}+27 m_1 m_3 r^7+54 m_2
   r^{15}+90 m_2 r^{13}-147 m_2 r^{12}-117 m_2 r^{10}+96
   m_2 r^9-3 m_2 r^6+27 m_2 r^4-9 m_3^2 r^{10}-3
   m_4 r^{15}+45 m_4 r^{13}-3 m_4 r^{12}-81 m_4
   r^{10}+6 m_4 r^9+36 m_4 r^7+8 r^{20}-11 r^{17}-12
   r^{15}-24 r^{14}-45 r^{13}+72 r^{12}+49 r^{11}+144 r^{10}-126
   r^9-22 r^8-171 r^7+84 r^6+90 r^4-18 r^3-18 r
$

$ 
a_2(r) \equiv 675 m_0 m_1 r^8-513 m_0 m_1 r^5+54 m_0 m_1
   r^2+135 m_0 m_3 r^8-81 m_0 m_3 r^5-360 m_1
   r^{13}+777 m_1 r^{10}-675 m_1 r^8-213 m_1 r^7+513
   m_1 r^5+12 m_1 r^4-54 m_1 r^2-60 m_3 r^{13}+111
   m_3 r^{10}-135 m_3 r^8+3 m_3 r^7+81 m_3 r^5
$

$  
a_3(r) \equiv -360 m_0^2 r^6+315 m_0^2 r^3-36 m_0^2+420 m_0
   r^{11}-1080 m_0 r^8+720 m_0 r^6+522 m_0 r^5-630
   m_0 r^3-24 m_0 r^2+72 m_0+180 r^{13}-420 r^{11}-252
   r^{10}+1080 r^8-9 r^7-360 r^6-522 r^5+315 r^3+24 r^2-36
$.

\subsubsection{$p_2 \left(r,\vec{r} \cdot \vec{v} \right)$}
The condition $q^{(6)}(t_0)=m_6$ and the previous equations imply the following polynomial equation in $r$ and $\vec{r} \cdot \vec{v}$:

\begin{equation}
p_2\left(r,\vec{r} \cdot \vec{v}\right)=0	
\end{equation}
where
\begin{equation}
p_2\left(r,\vec{r} \cdot \vec{v}\right)=b_0(r)+b_1(r) \vec{r} \cdot \vec{v}+b_2(r)\left(\vec{r} \cdot \vec{v}\right)^2+b_3(r)\left(\vec{r} \cdot \vec{v}\right)^3+b_4(r)\left(\vec{r} \cdot \vec{v}\right)^4
\end{equation}
and

$  
b_0(r) \equiv -8 r^{29}+8 m_0 r^{27}-24 m_4 r^{27}-16 {m6} r^{27}-8
   r^{27}+27 r^{26}-39 m_0 r^{24}-60 m_2 r^{24}+28 m_4
   r^{24}+40 {m6} r^{24}+39 r^{24}-6 r^{23}+57 m_0^2 r^{22}+252
   m_2^2 r^{22}+9 m_4^2 r^{22}-114 m_0 r^{22}+216 m_0
   m_2 r^{22}-216 m_2 r^{22}+18 m_0 m_4 r^{22}+144
   m_2 m_4 r^{22}-18 m_4 r^{22}-24 m_0 {m6}
   r^{22}+24 {m6} r^{22}+57 r^{22}+102 m_0 r^{21}+314 m_2
   r^{21}+26 m_4 r^{21}-33 {m6} r^{21}-102 r^{21}-45 m_0^3
   r^{20}+135 m_0^2 r^{20}-180 m_0 m_2^2 r^{20}+180
   m_2^2 r^{20}-45 m_0 m_4^2 r^{20}+45 m_4^2 r^{20}-135
   m_0 r^{20}-180 m_0^2 m_2 r^{20}+360 m_0 m_2
   r^{20}-180 m_2 r^{20}-90 m_0^2 m_4 r^{20}+180 m_0
   m_4 r^{20}-180 m_0 m_2 m_4 r^{20}+180 m_2
   m_4 r^{20}-90 m_4 r^{20}-41 r^{20}-330 m_0^2 r^{19}-270
   m_2^2 r^{19}-18 m_4^2 r^{19}+660 m_0 r^{19}-882 m_0
   m_2 r^{19}+882 m_2 r^{19}-228 m_0 m_4 r^{19}-279
   m_2 m_4 r^{19}+228 m_4 r^{19}+54 m_0 {m6}
   r^{19}-54 {m6} r^{19}-330 r^{19}-196 m_0 r^{18}-626 m_2
   r^{18}-51 m_4 r^{18}+10 {m6} r^{18}+196 r^{18}+234 m_0^3
   r^{17}-702 m_0^2 r^{17}+162 m_0 m_2^2 r^{17}-162
   m_2^2 r^{17}+90 m_0 m_4^2 r^{17}-90 m_4^2 r^{17}+702
   m_0 r^{17}+540 m_0^2 m_2 r^{17}-1080 m_0 m_2
   r^{17}+540 m_2 r^{17}+306 m_0^2 m_4 r^{17}-612 m_0
   m_4 r^{17}+261 m_0 m_2 m_4 r^{17}-261 m_2
   m_4 r^{17}+306 m_4 r^{17}-9 m_0^2 {m6} r^{17}+18
   m_0 {m6} r^{17}-9 {m6} r^{17}-90 r^{17}+786 m_0^2
   r^{16}-234 m_2^2 r^{16}+9 m_4^2 r^{16}-1572 m_0
   r^{16}+1236 m_0 m_2 r^{16}-1236 m_2 r^{16}+486 m_0
   m_4 r^{16}+126 m_2 m_4 r^{16}-486 m_4 r^{16}-36
   m_0 {m6} r^{16}+36 {m6} r^{16}+786 r^{16}+264 m_0
   r^{15}+558 m_2 r^{15}+32 m_4 r^{15}-{m6} r^{15}-264
   r^{15}-504 m_0^3 r^{14}+1512 m_0^2 r^{14}+198 m_0
   m_2^2 r^{14}-198 m_2^2 r^{14}-45 m_0 m_4^2 r^{14}+45
   m_4^2 r^{14}-1512 m_0 r^{14}-522 m_0^2 m_2
   r^{14}+1044 m_0 m_2 r^{14}-522 m_2 r^{14}-423 m_0^2
   m_4 r^{14}+846 m_0 m_4 r^{14}+18 m_0 m_2 m_4
   r^{14}-18 m_2 m_4 r^{14}-423 m_4 r^{14}+18 m_0^2
   {m6} r^{14}-36 m_0 {m6} r^{14}+18 {m6} r^{14}+411
   r^{14}-984 m_0^2 r^{13}+270 m_2^2 r^{13}+1968 m_0
   r^{13}-648 m_0 m_2 r^{13}+648 m_2 r^{13}-360 m_0
   m_4 r^{13}+9 m_2 m_4 r^{13}+360 m_4 r^{13}+6
   m_0 {m6} r^{13}-6 {m6} r^{13}-984 r^{13}-219 m_0
   r^{12}-194 m_2 r^{12}-11 m_4 r^{12}+219 r^{12}+576 m_0^3
   r^{11}-1728 m_0^2 r^{11}-162 m_0 m_2^2 r^{11}+162
   m_2^2 r^{11}+1728 m_0 r^{11}+90 m_0^2 m_2 r^{11}-180
   m_0 m_2 r^{11}+90 m_2 r^{11}+288 m_0^2 m_4
   r^{11}-576 m_0 m_4 r^{11}-99 m_0 m_2 m_4
   r^{11}+99 m_2 m_4 r^{11}+288 m_4 r^{11}-9 m_0^2
   {m6} r^{11}+18 m_0 {m6} r^{11}-9 {m6} r^{11}-554
   r^{11}+681 m_0^2 r^{10}-18 m_2^2 r^{10}-1362 m_0
   r^{10}+36 m_0 m_2 r^{10}-36 m_2 r^{10}+84 m_0
   m_4 r^{10}-84 m_4 r^{10}+681 r^{10}+98 m_0 r^9+8
   m_2 r^9-98 r^9-369 m_0^3 r^8+1107 m_0^2 r^8-18 m_0
   m_2^2 r^8+18 m_2^2 r^8-1107 m_0 r^8+126 m_0^2
   m_2 r^8-252 m_0 m_2 r^8+126 m_2 r^8-81 m_0^2
   m_4 r^8+162 m_0 m_4 r^8-81 m_4 r^8+369 r^8-246
   m_0^2 r^7+492 m_0 r^7+42 m_0 m_2 r^7-42 m_2
   r^7-246 r^7-18 m_0 r^6+18 r^6+126 m_0^3 r^5-378 m_0^2
   r^5+378 m_0 r^5-54 m_0^2 m_2 r^5+108 m_0 m_2
   r^5-54 m_2 r^5-126 r^5+36 m_0^2 r^4-72 m_0 r^4+36
   r^4-18 m_0^3 r^2+54 m_0^2 r^2-54 m_0 r^2+18 r^2
$

$  
b_1(r) \equiv 900 m_1 r^{22}+492 m_3 r^{22}-1260 m_0 m_1 r^{20}+1260
   m_1 r^{20}-2520 m_1 m_2 r^{20}-180 m_0 m_3
   r^{20}-360 m_2 m_3 r^{20}+180 m_3 r^{20}-1260 m_1
   m_4 r^{20}-180 m_3 m_4 r^{20}-3927 m_1 r^{19}-1437
   m_3 r^{19}+4230 m_0 m_1 r^{17}-4230 m_1 r^{17}-450
   m_1 m_2 r^{17}+1026 m_0 m_3 r^{17}-342 m_2
   m_3 r^{17}-1026 m_3 r^{17}+1845 m_1 m_4 r^{17}+171
   m_3 m_4 r^{17}+5997 m_1 r^{16}+1440 m_3 r^{16}+405
   m_0^2 m_1 r^{15}-810 m_0 m_1 r^{15}+405 m_1
   r^{15}+810 m_0 m_1 m_2 r^{15}-810 m_1 m_2
   r^{15}+135 m_0^2 m_3 r^{15}-270 m_0 m_3 r^{15}+270
   m_0 m_2 m_3 r^{15}-270 m_2 m_3 r^{15}+135
   m_3 r^{15}+405 m_0 m_1 m_4 r^{15}-405 m_1
   m_4 r^{15}+135 m_0 m_3 m_4 r^{15}-135 m_3
   m_4 r^{15}-5526 m_0 m_1 r^{14}+5526 m_1 r^{14}+4707
   m_1 m_2 r^{14}-1620 m_0 m_3 r^{14}+765 m_2
   m_3 r^{14}+1620 m_3 r^{14}-666 m_1 m_4 r^{14}+9
   m_3 m_4 r^{14}-3915 m_1 r^{13}-537 m_3 r^{13}-675
   m_0^2 m_1 r^{12}+1350 m_0 m_1 r^{12}-675 m_1
   r^{12}-945 m_0 m_1 m_2 r^{12}+945 m_1 m_2
   r^{12}-108 m_0^2 m_3 r^{12}+216 m_0 m_3 r^{12}-243
   m_0 m_2 m_3 r^{12}+243 m_2 m_3 r^{12}-108
   m_3 r^{12}-270 m_0 m_1 m_4 r^{12}+270 m_1
   m_4 r^{12}-135 m_0 m_3 m_4 r^{12}+135 m_3
   m_4 r^{12}+3690 m_0 m_1 r^{11}-3690 m_1 r^{11}-1980
   m_1 m_2 r^{11}+882 m_0 m_3 r^{11}-63 m_2 m_3
   r^{11}-882 m_3 r^{11}+81 m_1 m_4 r^{11}+1047 m_1
   r^{10}+42 m_3 r^{10}-27 m_0^2 m_1 r^9+54 m_0
   m_1 r^9-27 m_1 r^9+648 m_0 m_1 m_2 r^9-648
   m_1 m_2 r^9-189 m_0^2 m_3 r^9+378 m_0 m_3
   r^9-27 m_0 m_2 m_3 r^9+27 m_2 m_3 r^9-189
   m_3 r^9-135 m_0 m_1 m_4 r^9+135 m_1 m_4
   r^9-1422 m_0 m_1 r^8+1422 m_1 r^8+243 m_1 m_2
   r^8-108 m_0 m_3 r^8+108 m_3 r^8-102 m_1 r^7+459
   m_0^2 m_1 r^6-918 m_0 m_1 r^6+459 m_1 r^6-513
   m_0 m_1 m_2 r^6+513 m_1 m_2 r^6+162 m_0^2
   m_3 r^6-324 m_0 m_3 r^6+162 m_3 r^6+288 m_0
   m_1 r^5-288 m_1 r^5-162 m_0^2 m_1 r^3+324 m_0
   m_1 r^3-162 m_1 r^3
$

$
b_2(r) \equiv -576 r^{22}-1080 m_0 r^{20}-1440 m_2 r^{20}+360 m_4
   r^{20}+1080 r^{20}+2466 r^{19}+2520 m_0^2 r^{18}-5040 m_0
   r^{18}+5040 m_0 m_2 r^{18}-5040 m_2 r^{18}+2520 m_0
   m_4 r^{18}-2520 m_4 r^{18}+2520 r^{18}+6156 m_0
   r^{17}+6876 m_2 r^{17}-108 m_4 r^{17}-6156 r^{17}-3375
   r^{16}-10170 m_0^2 r^{15}+18900 m_1^2 r^{15}+540 m_3^2
   r^{15}+20340 m_0 r^{15}-7380 m_0 m_2 r^{15}+7380 m_2
   r^{15}+6480 m_1 m_3 r^{15}-5400 m_0 m_4 r^{15}+5400
   m_4 r^{15}-10170 r^{15}-11259 m_0 r^{14}-6480 m_2
   r^{14}-297 m_4 r^{14}+11259 r^{14}+540 m_0^3 r^{13}-1620
   m_0^2 r^{13}+1620 m_0 r^{13}+1080 m_0^2 m_2
   r^{13}-2160 m_0 m_2 r^{13}+1080 m_2 r^{13}+540 m_0^2
   m_4 r^{13}-1080 m_0 m_4 r^{13}+540 m_4 r^{13}+1116
   r^{13}+16911 m_0^2 r^{12}-14580 m_1^2 r^{12}-54 m_3^2
   r^{12}-33822 m_0 r^{12}+1350 m_0 m_2 r^{12}-1350 m_2
   r^{12}-2970 m_1 m_3 r^{12}+3402 m_0 m_4 r^{12}-3402
   m_4 r^{12}+16911 r^{12}+8694 m_0 r^{11}+1089 m_2
   r^{11}+45 m_4 r^{11}-8694 r^{11}-2565 m_0^3 r^{10}+7695
   m_0^2 r^{10}+4050 m_0 m_1^2 r^{10}-4050 m_1^2
   r^{10}-7695 m_0 r^{10}-1890 m_0^2 m_2 r^{10}+3780 m_0
   m_2 r^{10}-1890 m_2 r^{10}+810 m_0 m_1 m_3
   r^{10}-810 m_1 m_3 r^{10}-1485 m_0^2 m_4 r^{10}+2970
   m_0 m_4 r^{10}-1485 m_4 r^{10}+2394 r^{10}-14508
   m_0^2 r^9+3780 m_1^2 r^9+29016 m_0 r^9+1386 m_0
   m_2 r^9-1386 m_2 r^9+378 m_1 m_3 r^9-522 m_0
   m_4 r^9+522 m_4 r^9-14508 r^9-2835 m_0 r^8-45 m_2
   r^8+2835 r^8+4374 m_0^3 r^7-13122 m_0^2 r^7-4860 m_0
   m_1^2 r^7+4860 m_1^2 r^7+13122 m_0 r^7-351 m_0^2
   m_2 r^7+702 m_0 m_2 r^7-351 m_2 r^7-810 m_0
   m_1 m_3 r^7+810 m_1 m_3 r^7+945 m_0^2 m_4
   r^7-1890 m_0 m_4 r^7+945 m_4 r^7-4374 r^7+6363
   m_0^2 r^6-324 m_1^2 r^6-12726 m_0 r^6-396 m_0
   m_2 r^6+396 m_2 r^6+6363 r^6+324 m_0 r^5-324 r^5-3213
   m_0^3 r^4+9639 m_0^2 r^4+810 m_0 m_1^2 r^4-810
   m_1^2 r^4-9639 m_0 r^4+1161 m_0^2 m_2 r^4-2322
   m_0 m_2 r^4+1161 m_2 r^4+3213 r^4-1116 m_0^2
   r^3+2232 m_0 r^3-1116 r^3+864 m_0^3 r-2592 m_0^2
   r+2592 m_0 r-864 r
$

$
b_3(r) \equiv -15660 m_0^2 m_1 r^8+18090 m_0^2 m_1 r^5-2430 m_0^2
   m_1 r^2-2700 m_0^2 m_3 r^8+2700 m_0^2 m_3
   r^5-39060 m_0 m_1 r^{13}+33030 m_0 m_1 r^{10}+31320
   m_0 m_1 r^8-6120 m_0 m_1 r^7-36180 m_0 m_1
   r^5+486 m_0 m_1 r^4+4860 m_0 m_1 r^2-7740 m_0
   m_3 r^{13}+6030 m_0 m_3 r^{10}+5400 m_0 m_3
   r^8-1206 m_0 m_3 r^7-5400 m_0 m_3 r^5+11760 m_1
   r^{18}-39420 m_1 r^{15}+39060 m_1 r^{13}+19170 m_1
   r^{12}-33030 m_1 r^{10}-3354 m_1 r^9-15660 m_1 r^8+6120
   m_1 r^7+180 m_1 r^6+18090 m_1 r^5-486 m_1 r^4-2430
   m_1 r^2+1680 m_3 r^{18}-5220 m_3 r^{15}+7740 m_3
   r^{13}+594 m_3 r^{12}-6030 m_3 r^{10}+30 m_3 r^9-2700
   m_3 r^8+1206 m_3 r^7+2700 m_3 r^5
$

$
b_4(r) \equiv 9720 m_0^3 r^6-11340 m_0^3 r^3+1620 m_0^3+15120 m_0^2
   r^{11}-2700 m_0^2 r^8-29160 m_0^2 r^6-8370 m_0^2
   r^5+34020 m_0^2 r^3+324 m_0^2 r^2-4860 m_0^2-15120
   m_0 r^{16}+53280 m_0 r^{13}-30240 m_0 r^{11}-37260
   m_0 r^{10}+5400 m_0 r^8+8208 m_0 r^7+29160 m_0
   r^6+16740 m_0 r^5-360 m_0 r^4-34020 m_0 r^3-648
   m_0 r^2+4860 m_0-5040 r^{18}+15120 r^{16}+10800 r^{15}-53280
   r^{13}-1296 r^{12}+15120 r^{11}+37260 r^{10}-90 r^9-2700 r^8-8208
   r^7-9720 r^6-8370 r^5+360 r^4+11340 r^3+324 r^2-1620
 $.

\subsubsection{Finding candidate values of $r$ and $\vec{r} \cdot \vec{v}$}
Solving the equations $p_1 \left(r,\vec{r} \cdot \vec{v} \right)=0$
and $p_2 \left(r,\vec{r} \cdot \vec{v} \right)=0$ simultaneously,
we get a fairly short list of possible solution pairs 
$\left(r,\vec{r} \cdot \vec{v} \right).$ Some of them can be discarded immediately:

\begin{enumerate}
	\item Pairs for which $r$ is not real;
	\item Pairs for which $r \leqslant 0$;
	\item Pairs for which $\vec{r} \cdot \vec{v}$ is not real.
\end{enumerate}

For each of the remaining $\left(r,\vec{r} \cdot \vec{v} \right)$ pairs, we then perform the following:

\begin{enumerate}
	\item Compute $\vec{v} \cdot \vec{v}$ using an equation presented earlier, and discard the pair if this result is not positive;
	\item Compute $\left| \vec{v} \right|=\sqrt{\vec{v} \cdot \vec{v}}$;
	\item Discard the pair if the inequality $\left| \vec{r} \cdot \vec{v} \right| \leqslant r \left| \vec{v} \right|$ is not satisfied.
\end{enumerate}

We are then in a position to compute, for each remaining pair, its accompanying values of $\vec{r} \cdot \hat{R}$, 
$\vec{r} \cdot \hat{T}$,
$\vec{v} \cdot \hat{R}$, and
$\vec{v} \cdot \hat{T}$.

All that now remains is to determine for each pair the possibilities for $\vec{r} \cdot \hat{H}$ and
$\vec{v} \cdot \hat{H}$.

For the simultaneous equations

\begin{equation}
\begin{split}
\vec{r} \cdot \hat{H}^2+\vec{r} \cdot \hat{R}^2+\vec{r} \cdot \hat{T}^2=r^2\\
   \vec{v} \cdot \hat{H}^2+\vec{v} \cdot \hat{R}^2+\vec{v} \cdot \hat{T}^2=v^2\\
    \vec{r} \cdot \hat{H} \vec{v} \cdot \hat{H}+\vec{r} \cdot \hat{R}
   \vec{v} \cdot \hat{R}+\vec{r} \cdot \hat{T} \vec{v} \cdot \hat{T}=\vec{r} \cdot \vec{v}
\end{split}
\end{equation}
to have a solution, we must have

\begin{equation}
\begin{split}
\left| \vec{r} \cdot \vec{v}-\vec{r} \cdot \hat{R} \, \vec{v} \cdot \hat{R}-\vec{r} \cdot \hat{T} \, \vec{v} \cdot \hat{T} \right|\\=\sqrt{r^2-\left( \vec{r} \cdot \hat{R} \right)^2-\left( \vec{r} \cdot \hat{T} \right)^2}
   \sqrt{v^2-\left( \vec{v} \cdot \hat{R} \right)^2-\left( \vec{v} \cdot \hat{T} \right)^2}	
   \end{split}
\end{equation}

We therefore discard pairs not satisfying this condition.

\subsection{Case $r = 1$, $\vec{r} \cdot \vec{v}=0$, $\vec{r} \neq \hat{R}$}
 Solving the system

\begin{equation}
\begin{array}{c}
  q(t_0)=m_0 \\
  q'(t_0)=m_1 \\
  q''(t_0)=m_2 \\
  q'''(t_0)=m_3 \\
  q^{(4)}(t_0)=m_3 \\
  r=1\\
  \vec{r} \cdot \vec{v}=0
\end{array}
\end{equation}

we get

 \begin{equation}
 \begin{split}
 	\vec{r} \cdot \hat{R}=\frac{2-m_0}{2}\\
   \vec{v} \cdot \hat{R}=-\frac{2 (16 m_0+5 (4 m_2+m_4))
   \vec{r} \cdot \hat{T}+m_0 m_5}{32 m_0+6 (4
   m_2+m_4)}\\
   \vec{v} \cdot \hat{T}=\frac{12 m_0 - 6 m_0^2 + 8 m_2 - 3 m_0 m_2 + 2 m_4}{12 m_0}\\
   \vec{v} \cdot \vec{v}=\frac{3 m_0+4 m_2+m_4}{3
   m_0}\\
 \end{split}
 \end{equation}
 
\subsection{Case $r = 1$, $\vec{r} \cdot \vec{v} \neq 0$, $\vec{r} \neq \hat{R}$}
 Solving the system

\begin{equation}
\begin{array}{c}
  q(t_0)=m_0 \\
  q'(t_0)=m_1 \\
  q''(t_0)=m_2 \\
  q'''(t_0)=m_3 \\
  q^{(4)}(t_0)=m_3 \\
\end{array}
\end{equation}

we get two possible solutions:

\begin{equation}
   \vec{r} \cdot \hat{R}= 1-\frac{m_0}{2} \\ 
\end{equation}
\begin{equation}
   \begin{split}
   \vec{r} \cdot \hat{T}= -\frac{3 m_4 m_0^2}{2 (48
   m_0 m_1+12 m_0 m_3)}-\frac{6 m_2 m_0^2}{48 m_0 m_1+12 m_0 m_3}\\
   -\frac{66
   m_1^2 m_0}{48 m_0 m_1+12 m_0 m_3}
   -\frac{15 m_3^2 m_0}{2 (48 m_0 m_1+12
   m_0 m_3)}\\
   -\frac{93 m_1 m_3 m_0}{2 (48 m_0 m_1+12 m_0
   m_3)}
   -\frac{m_1}{2}
   +\frac{160 m_1^2}{48 m_0 m_1+12 m_0 m_3}\\
   +\frac{10 m_3^2}{48
   m_0 m_1+12 m_0 m_3}
   +\frac{80 m_1 m_3}{48 m_0 m_1+12 m_0 m_3}
   +\frac{4
   m_1}{3 m_0}
   +\frac{m_3}{3 m_0}\\
   -\frac{\sqrt{\gamma }}{18 (48 m_0 m_1+12 m_0 m_3)
   m_0^2} \\ 
   \end{split}
\end{equation}
\begin{equation}
   \vec{r} \cdot \vec{v}= \frac{4 m_1}{3 m_0}+\frac{m_3}{3 m_0} \\ 
   \end{equation}
   \begin{equation}
   \begin{split}
   \vec{v} \cdot \hat{R}=
   \frac{3 m_4 m_0^2}{2 (48 m_0 m_1+12 m_0 m_3)}
   +\frac{6 m_2 m_0^2}{48 m_0
   m_1+12 m_0 m_3}\\
   +\frac{66 m_1^2 m_0}{48 m_0 m_1+12 m_0 m_3}
   +\frac{15 m_3^2
   m_0}{2 (48 m_0 m_1+12 m_0 m_3)}\\
   +\frac{93 m_1 m_3 m_0}{2 (48 m_0 m_1+12
   m_0 m_3)}
   -\frac{160 m_1^2}{48 m_0 m_1+12 m_0 m_3}\\
   -\frac{10 m_3^2}{48 m_0
   m_1+12 m_0 m_3}
   -\frac{80 m_1 m_3}{48 m_0 m_1+12 m_0 m_3}\\
   +\frac{\sqrt{\gamma }}{18
   (48 m_0 m_1+12 m_0 m_3) m_0^2} \\ 
   \end{split}
\end{equation}
\begin{equation}
   \begin{split}
   \vec{v} \cdot \hat{T}= -\frac{17 m_1^2}{9
   m_0^2}
   +\frac{16 m_1^2}{3 m_0^3}
   -\frac{37 m_3 m_1}{36 m_0^2}
   +\frac{8 m_3 m_1}{3
   m_0^3}
   -\frac{5 m_3^2}{36 m_0^2}\\
   +\frac{m_3^2}{3 m_0^3}
   -\frac{m_0}{2}+\frac{m_2}{3
   m_0}-\frac{m_2}{4}
   +\frac{m_4}{12 m_0}
   -\frac{\sqrt{\gamma }}{324 m_0^5}+1 \\ 
   \end{split}
\end{equation}
\begin{equation}
   \begin{split}
   \vec{v} \cdot \vec{v}=
   -\frac{34 m_1^2}{9 m_0^2}+\frac{32 m_1^2}{3 m_0^3}
   -\frac{37 m_3 m_1}{18 m_0^2}
   +\frac{16
   m_3 m_1}{3 m_0^3}-\frac{5 m_3^2}{18 m_0^2}\\
   +\frac{2 m_3^2}{3 m_0^3}+\frac{2 m_2}{3
   m_0}+\frac{m_4}{6 m_0}
   -\frac{\sqrt{\gamma }}{162 m_0^5}+1 \\ 
   \end{split}
\end{equation}
\begin{equation}
   \vec{r} \cdot \hat{R}=
   1-\frac{m_0}{2} \\ 
   \end{equation}
   
   or
   
\begin{equation}
   \begin{split}\vec{r} \cdot \hat{T}= -\frac{3 m_4 m_0^2}{2 (48 m_0 m_1+12 m_0
   m_3)}-\frac{6 m_2 m_0^2}{48 m_0 m_1+12 m_0 m_3}\\
   -\frac{66 m_1^2 m_0}{48 m_0
   m_1+12 m_0 m_3}
   -\frac{15 m_3^2 m_0}{2 (48 m_0 m_1+12 m_0 m_3)}\\
   -\frac{93 m_1
   m_3 m_0}{2 (48 m_0 m_1+12 m_0 m_3)}-\frac{m_1}{2}
   +\frac{160 m_1^2}{48 m_0
   m_1+12 m_0 m_3}\\
   +\frac{10 m_3^2}{48 m_0 m_1+12 m_0 m_3}
   +\frac{80 m_1 m_3}{48
   m_0 m_1+12 m_0 m_3}
   +\frac{4 m_1}{3 m_0}+\frac{m_3}{3 m_0}\\
   +\frac{\sqrt{\gamma }}{18 (48
   m_0 m_1+12 m_0 m_3) m_0^2} \\ 
   \end{split}
\end{equation}
\begin{equation}
   \vec{r} \cdot \vec{v}= \frac{4 m_1}{3
   m_0}+\frac{m_3}{3 m_0} \\ 
\end{equation}
\begin{equation}
   \begin{split}
   \vec{v} \cdot \hat{R}= \frac{3 m_4 m_0^2}{2 (48 m_0 m_1+12
   m_0 m_3)}
   +\frac{6 m_2 m_0^2}{48 m_0 m_1+12 m_0 m_3}\\
   +\frac{66 m_1^2 m_0}{48
   m_0 m_1+12 m_0 m_3}
   +\frac{15 m_3^2 m_0}{2 (48 m_0 m_1+12 m_0 m_3)}\\
   +\frac{93
   m_1 m_3 m_0}{2 (48 m_0 m_1+12 m_0 m_3)}
   -\frac{160 m_1^2}{48 m_0 m_1+12
   m_0 m_3}\\
   -\frac{10 m_3^2}{48 m_0 m_1+12 m_0 m_3}
   -\frac{80 m_1 m_3}{48 m_0
   m_1+12 m_0 m_3}\\
   -\frac{\sqrt{\gamma }}{18 (48 m_0 m_1+12 m_0 m_3)
   m_0^2} \\ 
   \end{split}
\end{equation}
\begin{equation}
   \begin{split}
   \vec{v} \cdot \hat{T}= -\frac{17 m_1^2}{9 m_0^2}+\frac{16 m_1^2}{3 m_0^3}-\frac{37 m_3
   m_1}{36 m_0^2}+\frac{8 m_3 m_1}{3 m_0^3}-\frac{5 m_3^2}{36 m_0^2}\\
   +\frac{m_3^2}{3
   m_0^3}
   -\frac{m_0}{2}
   +\frac{m_2}{3 m_0}-\frac{m_2}{4}+\frac{m_4}{12 m_0}+\frac{\sqrt{\gamma
   }}{324 m_0^5}+1 \\ 
   \end{split}
\end{equation}
\begin{equation}
   \begin{split}\vec{v} \cdot \vec{v}= -\frac{34 m_1^2}{9 m_0^2}+\frac{32 m_1^2}{3 m_0^3}-\frac{37
   m_3 m_1}{18 m_0^2}+\frac{16 m_3 m_1}{3 m_0^3}-\frac{5 m_3^2}{18 m_0^2}\\
   +\frac{2
   m_3^2}{3 m_0^3}
   +\frac{2 m_2}{3 m_0}+\frac{m_4}{6 m_0}+\frac{\sqrt{\gamma }}{162
   m_0^5}+1
   \end{split}
\end{equation}

where

\begin{equation}
\gamma=\delta^2
   -324 m_0^5 \epsilon,	
\end{equation}

\begin{equation}
\begin{split}
\delta=-162 m_0^5-108 m_0^4 m_2-27 m_0^4 m_4+612 m_0^3 m_1^2
+333 m_0^3 m_1
   m_3\\
   +45 m_0^3 m_3^2-1728 m_0^2 m_1^2
   -864 m_0^2 m_1 m_3
   -108 m_0^2
   m_3^2,
\end{split}
\end{equation}
and
\begin{equation}
\begin{split}
\epsilon=81 m_0^5+108 m_0^4 m_2+27 m_0^4 m_4-468 m_0^3
   m_1^2-117 m_0^3 m_1 m_3\\+36 m_0^3 m_1 m_5+9 m_0^3 m_3 m_5+240 m_0^2
   m_1^2 m_2+240 m_0^2 m_1^2 m_4\\
   +1728 m_0^2 m_1^2+120 m_0^2 m_1 m_2
   m_3+120 m_0^2 m_1 m_3 m_4\\
   +864 m_0^2 m_1 m_3+15 m_0^2 m_2 m_3^2+15
   m_0^2 m_3^2 m_4+108 m_0^2 m_3^2\\
   -1600 m_0 m_1^4-1840 m_0 m_1^3 m_3-780
   m_0 m_1^2 m_3^2-145 m_0 m_1 m_3^3\\
   -10 m_0 m_3^4+2560 m_1^4
   +2560 m_1^3
   m_3+960 m_1^2 m_3^2+160 m_1 m_3^3+10 m_3^4.
\end{split}
\end{equation}

\section{Operational Strategy}
We have identified three possible exhaustive and mutually exclusive cases:
\begin{enumerate}
	\item $r \neq 1;$
	\item $r = 1$, $\vec{r} \cdot \vec{v}=0$, $\vec{r} \neq \hat{R};$ and
	\item $r = 1$, $\vec{r} \cdot \vec{v} \neq 0$, $\vec{r} \neq \hat{R}.$
\end{enumerate}
Naturally, when processing data the applicable case is unknown; that is something we are trying to determine. Therefore three data processing procedures should be conducted in parallel, treating the incoming data under the assumptions 

$r \neq 1;$ 

$r = 1$, $\vec{r} \cdot \vec{v}=0$, $\vec{r} \neq \hat{R};$ and
	
$r = 1$, $\vec{r} \cdot \vec{v} \neq 0$, $\vec{r} \neq \hat{R},$ respectively.

The orbital elements produced by the case actually in force will converge rapidly, while the other two procedures will fail to converge, since they are based on assumptions not in force. In practice, it will rarely (if ever) happen that $r$ is precisely equal to $1,$ so the second and third cases are very unlikely to be encountered.

\section{Example}

Suppose we have data as follows at some time $t_0$:

\begin{equation}
\begin{split}
m_0=2.73407\\
m_1=2.90282\\
m_2=-1.37835\\
m_3=-3.62096\\
m_4=8.01712\\
m_5=-14.8261\\
m_6=90.5907.
\end{split}
\end{equation}

We then solve simultaneously (see Appendix for one algorithm), the equations

\noindent\(8 r^{20} \vec{r} \cdot \vec{v}-11 r^{17} \vec{r} \cdot \vec{v}-77.6737 r^{15} \vec{r} \cdot \vec{v}-24 r^{14} \vec{r} \cdot \vec{v}+180 r^{13}
   \left(\vec{r} \cdot \vec{v}\right)^3-827.758 r^{13} \left(\vec{r} \cdot \vec{v}\right)^2-545.801 r^{13} \vec{r} \cdot \vec{v}+53.7139 r^{12} \vec{r} \cdot \vec{v}+728.309
   r^{11} \left(\vec{r} \cdot \vec{v}\right)^3+49 r^{11} \vec{r} \cdot \vec{v}-252 r^{10} \left(\vec{r} \cdot \vec{v}\right)^3+1853.56 r^{10} \left(\vec{r} \cdot \vec{v}\right)^2-73.0933
   r^{10} \vec{r} \cdot \vec{v}+134.273 r^9 \vec{r} \cdot \vec{v}-1872.79 r^8 \left(\vec{r} \cdot \vec{v}\right)^3+2550.08 r^8 \left(\vec{r} \cdot \vec{v}\right)^2-22 r^8
   \vec{r} \cdot \vec{v}-9 r^7 \left(\vec{r} \cdot \vec{v}\right)^3-629.164 r^7 \left(\vec{r} \cdot \vec{v}\right)^2-85.0766 r^7 \vec{r} \cdot \vec{v}-1082.52 r^6
   \left(\vec{r} \cdot \vec{v}\right)^3-141.527 r^6 \vec{r} \cdot \vec{v}+905.184 r^5 \left(\vec{r} \cdot \vec{v}\right)^3-2073.68 r^5 \left(\vec{r} \cdot \vec{v}\right)^2+34.8338 r^4
   \left(\vec{r} \cdot \vec{v}\right)^2+183.489 r^4 \vec{r} \cdot \vec{v}+947.203 r^3 \left(\vec{r} \cdot \vec{v}\right)^3+31.2132 r^3 \vec{r} \cdot \vec{v}-41.6176 r^2
   \left(\vec{r} \cdot \vec{v}\right)^3+271.819 r^2 \left(\vec{r} \cdot \vec{v}\right)^2-54.1259 r \vec{r} \cdot \vec{v}+76.6609 r^{20}-161.685 r^{17}+389.188
   r^{15}+127.533 r^{14}-800.469 r^{12}-91.169 r^{11}+508.491 r^9+63.1742 r^8-112.311 r^6-14.5141 r^5+15.1011
   r^3-108.252 \left(\vec{r} \cdot \vec{v}\right)^3=0\)

and

\noindent\(-576 r^{22} \left(\vec{r} \cdot \vec{v}\right)^2+831.027 r^{22} \vec{r} \cdot \vec{v}+2998.2 r^{20} \left(\vec{r} \cdot \vec{v}\right)^2-21023.9 r^{20}
   \vec{r} \cdot \vec{v}+2466 r^{19} \left(\vec{r} \cdot \vec{v}\right)^2-6196.06 r^{19} \vec{r} \cdot \vec{v}-5040 r^{18} \left(\vec{r} \cdot \vec{v}\right)^4+28054. r^{18}
   \left(\vec{r} \cdot \vec{v}\right)^3+30564.8 r^{18} \left(\vec{r} \cdot \vec{v}\right)^2+331.509 r^{17} \left(\vec{r} \cdot \vec{v}\right)^2+52917.1 r^{17} \vec{r} \cdot \vec{v}-26219.1
   r^{16} \left(\vec{r} \cdot \vec{v}\right)^4-3375 r^{16} \left(\vec{r} \cdot \vec{v}\right)^2+12194. r^{16} \vec{r} \cdot \vec{v}+10800 r^{15}
   \left(\vec{r} \cdot \vec{v}\right)^4-95527.8 r^{15} \left(\vec{r} \cdot \vec{v}\right)^3+10213.4 r^{15} \left(\vec{r} \cdot \vec{v}\right)^2+8330.29 r^{15} \vec{r} \cdot \vec{v}-12973.2
   r^{14} \left(\vec{r} \cdot \vec{v}\right)^2-48420. r^{14} \vec{r} \cdot \vec{v}+92391.2 r^{13} \left(\vec{r} \cdot \vec{v}\right)^4-148017. r^{13}
   \left(\vec{r} \cdot \vec{v}\right)^3+13013.5 r^{13} \left(\vec{r} \cdot \vec{v}\right)^2-9420.09 r^{13} \vec{r} \cdot \vec{v}-1296 r^{12} \left(\vec{r} \cdot \vec{v}\right)^4+53496.2
   r^{12} \left(\vec{r} \cdot \vec{v}\right)^3+2573.19 r^{12} \left(\vec{r} \cdot \vec{v}\right)^2-4362.68 r^{12} \vec{r} \cdot \vec{v}+45465.7 r^{11}
   \left(\vec{r} \cdot \vec{v}\right)^4+13935.7 r^{11} \left(\vec{r} \cdot \vec{v}\right)^2+22529.1 r^{11} \vec{r} \cdot \vec{v}-64611.4 r^{10} \left(\vec{r} \cdot \vec{v}\right)^4+128400.
   r^{10} \left(\vec{r} \cdot \vec{v}\right)^3+2902.67 r^{10} \left(\vec{r} \cdot \vec{v}\right)^2+2887.17 r^{10} \vec{r} \cdot \vec{v}-90 r^9
   \left(\vec{r} \cdot \vec{v}\right)^4-9844.69 r^9 \left(\vec{r} \cdot \vec{v}\right)^3-26316.7 r^9 \left(\vec{r} \cdot \vec{v}\right)^2-8355.46 r^9 \vec{r} \cdot \vec{v}-8118.88 r^8
   \left(\vec{r} \cdot \vec{v}\right)^4-107294. r^8 \left(\vec{r} \cdot \vec{v}\right)^3-4854.06 r^8 \left(\vec{r} \cdot \vec{v}\right)^2-7452.04 r^8 \vec{r} \cdot \vec{v}+14233.2 r^7
   \left(\vec{r} \cdot \vec{v}\right)^4-23233.7 r^7 \left(\vec{r} \cdot \vec{v}\right)^3-9206.31 r^7 \left(\vec{r} \cdot \vec{v}\right)^2-296.088 r^7 \vec{r} \cdot \vec{v}+50683.3 r^6
   \left(\vec{r} \cdot \vec{v}\right)^4+522.508 r^6 \left(\vec{r} \cdot \vec{v}\right)^3+17349.9 r^6 \left(\vec{r} \cdot \vec{v}\right)^2+5801.91 r^6 \vec{r} \cdot \vec{v}-25168.5 r^5
   \left(\vec{r} \cdot \vec{v}\right)^4+128505. r^5 \left(\vec{r} \cdot \vec{v}\right)^3+561.838 r^5 \left(\vec{r} \cdot \vec{v}\right)^2+1449.7 r^5 \vec{r} \cdot \vec{v}-624.265 r^4
   \left(\vec{r} \cdot \vec{v}\right)^4+2446.37 r^4 \left(\vec{r} \cdot \vec{v}\right)^3-9730.02 r^4 \left(\vec{r} \cdot \vec{v}\right)^2-59130.5 r^3 \left(\vec{r} \cdot \vec{v}\right)^4-3355.8 r^3
   \left(\vec{r} \cdot \vec{v}\right)^2-1414.06 r^3 \vec{r} \cdot \vec{v}+974.266 r^2 \left(\vec{r} \cdot \vec{v}\right)^4-21210.9 r^2 \left(\vec{r} \cdot \vec{v}\right)^3+4505.18 r
   \left(\vec{r} \cdot \vec{v}\right)^2-8 r^{29}-1627.99 r^{27}+27 r^{26}+3863.18 r^{24}-6 r^{23}-4398.84 r^{22}-3036.98
   r^{21}-3903.59 r^{20}+7842.16 r^{19}+1020.01 r^{18}+9614.64 r^{17}-747.967 r^{16}-145.371 r^{15}-10559.8
   r^{14}-5058.81 r^{13}-200.548 r^{12}+8507.07 r^{11}+3095.31 r^{10}+158.912 r^9-4458.32 r^8-840.107
   r^7-31.2132 r^6+880.82 r^5+108.252 r^4-93.858 r^2+8447.22 \left(\vec{r} \cdot \vec{v}\right)^4=0\)

The real solutions with positive $r$ are given by

$$
\begin{array}{ccc}
\hline
& r & \vec{r}\cdot\vec{v} \\
\hline
 1 & 1.68599 & -2.12935  \\
 2 & 1.19671 & 1.09521  \\
 3 & 0.869413 & 0.8242  \\
 4 & 0.632199 & -0.259071  \\
\end{array}
$$

\subsection{Compute corresponding $\vec{v} \cdot \vec{v}$ values}

$$
\begin{array}{cccc}
\hline
& r & \vec{r}\cdot\vec{v}  & \vec{v}\cdot\vec{v} \\
\hline
 1 & 1.68599 & -2.12935 & 4.76251 \\
 2 & 1.19671 & 1.09521 & 1.02249 \\
 3 & 0.869413 & 0.8242 & 1.47683 \\
 4 & 0.632199 & -0.259071 & 1.85481 \\
\end{array}
$$

\subsection{Get rid of negative $\vec{v} \cdot \vec{v}$ values, if any}
$$
\begin{array}{cccc}
\hline
& r & \vec{r}\cdot\vec{v}  & \vec{v}\cdot\vec{v} \\
\hline
 1 & 1.68599 & -2.12935 & 4.76251 \\
 2 & 1.19671 & 1.09521 & 1.02249 \\
 3 & 0.869413 & 0.8242 & 1.47683 \\
 4 & 0.632199 & -0.259071 & 1.85481 \\
\end{array}
$$

\subsection{Get rid of pairs for which $\left| \vec{r} \cdot \vec{v} \right| > r \sqrt{\vec{v} \cdot \vec{v}}$, if any}

$$
\begin{array}{cccc}
\hline
& r & \vec{r}\cdot\vec{v}  & \vec{v}\cdot\vec{v} \\
\hline
 1 & 1.68599 & -2.12935 & 4.76251 \\
 2 & 1.19671 & 1.09521 & 1.02249 \\
 3 & 0.869413 & 0.8242 & 1.47683 \\
 4 & 0.632199 & -0.259071 & 1.85481 \\
\end{array}
$$

\subsection{Compute possible values of $\vec{r} \cdot \hat{H}$ and $\vec{v} \cdot \hat{H}$}
So far we have determined the following candidate solution data:
\scriptsize
$$
\begin{array}{ccccccccc}
\hline
& r & v & \vec{r}\cdot\hat{R}  & \vec{r}\cdot\hat{T} & \vec{v}\cdot\hat{R} & \vec{v}\cdot\hat{T} & \xi & \vec{r} \cdot \vec{v}\\
\hline
 1 & 1.68599 & 2.18232 & 0.554241 & -5.6706 & 2.08984 & 2.76423
& 1.78813 & -2.12935 \\
 2 & 1.19671 & 1.01118 & -0.150981 & 0.11657 & -0.47277 & 0.318484
& -0.32438 & 1.09521 \\
 3 & 0.869413 & 1.21525 & -0.489095 & -0.148376 & -0.478834 & -0.108766 & -0.411787 & 0.8242 \\
 4 & 0.632199 & 1.36192 & -0.667196 & 2.40054 & -4.11102 & -1.17276
& -0.654373 & -0.259071 \\
\end{array}
$$
\normalsize

We now determine the accompanying possible values of $\vec{r} \cdot \hat{H}$ and $\vec{v} \cdot \hat{H}$ and the corresponding state vectors:

$$
\begin{array}{ccccccc}
\hline
  & \vec{r}\cdot\hat{R}  & \vec{r}\cdot\hat{T} & \vec{r}\cdot\hat{H} & \vec{v}\cdot\hat{R} & \vec{v}\cdot\hat{T} & \vec{v}\cdot\hat{H} \\
\hline
 1 & -0.150981 & 0.11657 & -1.18141 & -0.47277 & 0.318484 & -0.835194 \\
 2 & -0.150981 & 0.11657 & 1.18141 & -0.47277 & 0.318484 & 0.835194 \\
\end{array}
$$
\section{Concluding Remark}
	In conclusion, we return to the claim in the Abstract: ``it is theoretically possible to narrow the possibilities for the tracker's state vector to at worst a few, and at best only two." 
	
	``at worst a few" has been established above. Regarding ``at best only two," it cannot be expected that the state vector will be uniquely determined by a single measurement of $q$ and its first six time derivatives, because orbit $A$ proceeding from a state vector of the form $\vec{r}=a \hat{R} + b \hat{T} +c \hat{H}, \,  \vec{v}=d \hat{R}+e \hat{T} +f\hat{H}$ will present the same function $q$ as orbit $B$ proceeding from a state vector of the form $\vec{r}=a \hat{R} + b \hat{T} -c \hat{H}, \,  \vec{v}=d \hat{R}+e \hat{T} -f\hat{H}$, and these orbits are distinct whenever $c \neq 0$ and $f \neq 0,$ i.e. whenever the trackee and the tracker are not in the same orbital plane. If they are not coplanar, the orbit of the trackee will be narrowed down to two definite possibilities.
\section{Appendix}
 Using $(2.1.2)$ and 
 solving 
\begin{equation}
p_1\left(r,\vec{r} \cdot \vec{v}\right)=0
\end{equation}
for $\vec{r} \cdot \vec{v}$ yields
\scriptsize
\begin{equation}
\begin{split}
	x_1(r)=-\frac{2 \sqrt[3]{2} a_1(r)}
	{3 \sqrt[3]{-27
   a_0(r) a_1(r)^2+3 \sqrt{3} \sqrt{-14 a_0(r) a_1(r)^5
   +27
   a_0(r)^2 a_1(r)^4
   +3 a_1(r)^6}+7 a_1(r)^3}}\\
   +\frac{\sqrt[3]{-27
   a_0(r) a_1(r)^2+3 \sqrt{3} \sqrt{-14 a_0(r) a_1(r)^5+27
   a_0(r)^2 a_1(r)^4+3 a_1(r)^6}+7 a_1(r)^3}}{3 \sqrt[3]{2}
   a_1(r)}\\
   -\frac{1}{3}
   \end{split}
\end{equation}
\begin{equation}
\begin{split}
	x_2(r)=\frac{\sqrt[3]{2} \left(1+i
   \sqrt{3}\right) a_1(r)}{3 \sqrt[3]{-27 a_0(r) a_1(r)^2+3 \sqrt{3}
   \sqrt{-14 a_0(r) a_1(r)^5+27 a_0(r)^2 a_1(r)^4+3
   a_1(r)^6}+7 a_1(r)^3}}\\
   -\frac{\left(1-i \sqrt{3}\right) \sqrt[3]{-27
   a_0(r) a_1(r)^2
   +3 \sqrt{3} \sqrt{-14 a_0(r) a_1(r)^5+27
   a_0(r)^2 a_1(r)^4+3 a_1(r)^6}+7 a_1(r)^3}}{6 \sqrt[3]{2}
   a_1(r)}\\
   -\frac{1}{3}
\end{split}
\end{equation}
\begin{equation}
\begin{split}
x_3(r)=\frac{\sqrt[3]{2} \left(1-i
   \sqrt{3}\right) a_1(r)}{3 \sqrt[3]{-27 a_0(r) a_1(r)^2+3 \sqrt{3}
   \sqrt{-14 a_0(r) a_1(r)^5+27 a_0(r)^2 a_1(r)^4+3
   a_1(r)^6}+7 a_1(r)^3}}\\
   -\frac{\left(1+i \sqrt{3}\right) \sqrt[3]{-27
   a_0(r) a_1(r)^2+3 \sqrt{3} \sqrt{-14 a_0(r) a_1(r)^5+27
   a_0(r)^2 a_1(r)^4+3 a_1(r)^6}+7 a_1(r)^3}}{6 \sqrt[3]{2}
   a_1(r)}\\
   -\frac{1}{3}
\end{split}
\end{equation}
\normalsize
We then get three functions
\begin{equation}
	q_1(r) \equiv p_2\left(r,x_1(r)\right)
\end{equation}
\begin{equation}
	q_2(r) \equiv p_2\left(r,x_2(r)\right)
\end{equation}
\begin{equation}
	q_3(r) \equiv p_2\left(r,x_3(r)\right)
\end{equation}
A candidate solution $r$ will be a positive zero of one or more of these equations. Recall the expression of the function $p_2$ given in $(2.1.3).$ If $q_i(r)=0,$ then the corresponding candidate value of $\vec{r} \cdot \vec{v}$ will be $x_i(r).$

\end{document}